# Design and Evaluation of a Collective IO Model for Loosely Coupled Petascale Programming

Zhao Zhang[+], Allan Espinosa[*], Kamil Iskra[#], Ioan Raicu[*], Ian Foster[#*+], Michael Wilde[#+]
[+]Computation Institute, University of Chicago & Argonne National Laboratory, USA
[*]Department of Computer Science, University of Chicago, IL, USA
[#]Mathematics and Computer Science Division, Argonne National Laboratory, Argonne IL, USA
zhaozhang@uchicago.edu, aespinosa@cs.uchicago.edu, iskra@mcs.anl.gov, iraicu@cs.uchicago.edu, {foster,wilde}@mcs.anl.gov

**Abstract**

*Loosely coupled programming is a powerful paradigm for rapidly creating higher-level applications from scientific programs on petascale systems, typically using scripting languages. This paradigm is a form of many-task computing (MTC) which focuses on the passing of data between programs as ordinary files rather than messages. While it has the significant benefits of decoupling producer and consumer and allowing existing application programs to be executed in parallel with no recoding, its typical implementation using shared file systems places a high performance burden on the overall system and on the user who will analyze and consume the downstream data. Previous efforts have achieved great speedups with loosely coupled programs, but have done so with careful manual tuning of all shared file system access. In this work, we evaluate a prototype collective IO model for file-based MTC. The model enables efficient and easy distribution of input data files to computing nodes and gathering of output results from them. It eliminates the need for such manual tuning and makes the programming of large-scale clusters using a loosely coupled model easier. Our approach, inspired by in-memory approaches to collective operations for parallel programming, builds on fast local file systems to provide high-speed local file caches for parallel scripts, uses a broadcast approach to handle distribution of common input data, and uses efficient scatter/gather and caching techniques for input and output. We describe the design of the prototype model, its implementation on the Blue Gene/P supercomputer, and present preliminary measurements of its performance on synthetic benchmarks and on a large-scale molecular dynamics application.*

## 1 Overview

We define "loosely coupled applications" as programs that involve the sequenced execution of other programs. In this programming model, programs exchange data via files; the application typically involves a large number of invocations, often of several different programs; and programs are typically feature a high degree of inter-task parallelism, enabled by data independence within the flow graph of files. Applications are typically written in scripting languages (Perl, Python, Tcl, and numerous "shells") [Ousterhout1998], which facilitate both the invocation of application programs and the passing and manipulation of files for program inputs and outputs. This style of programming is extensively employed in virtually every domain of science. For example, biologists run PERL scripts of BLAST and PFAM; neuroscientists run shell scripts of AIR, AFNI and FSL; physicists analyze collision data with scripts that execute analysis applications written in ROOT.

It is difficult to efficiently map this common and useful programming model onto computing clusters of rapidly increasing scale. We note that we are mainly concerned here with applications running on what we term "petascale-precursor" systems, where the sheer parallelism of the computing nodes of the system can easily overwhelm a traditional IO subsystem, and in particular, its shared file systems. As clusters have grown larger, to tens or, recently, hundreds of thousands of nodes, the IO strategies of loosely coupled applications have become both a performance bottleneck and a source of complexity. Significant manual effort is needed to scale application performance as cluster size grows.

The specific problem we address here is that as the number of nodes in large-scale clusters contending for shared resources grows large, the IO bandwidth, volume and/or file management transaction rate exceeds some aggregate capacity limit, bottlenecks arise and the system becomes unbalanced. Thus, CPU cycles are wasted because the IO subsystem cannot service the CPUs fast enough. (We are concerned here with applications with high enough IO-to-compute ratios for IO to become the primary obstacle to parallel speedup. Applications that do relatively little IO while computing for long periods typically perform well in loosely coupled settings without any change to their IO strategy.)

While petascale systems have massive shared IO subsystems, these subsystems often have vulnerabilities in handling file management transactions (e.g., creating and writing huge numbers of files at high rates) that are ill-matched with the needs of loosely coupled programs. Our work remedies this deficiency and makes petascale systems *attractive* for this important and productive paradigm for knitting existing scientific programs into powerful workflows.

Our strategy of collective IO is inspired by the collective data operations employed by tightly coupled message passing programming models. In these models, data is exchanged, both between in-memory tasks and between tasks and files, using operations such as *scatter* (often assisted by *broadcast*) and *gather*. In our model:

- Input files are broadcast from shared file systems to local file systems.

- Output files are locally batched up from applications and efficiently transferred to shared persistent storage.
- Intermediate file systems are provided within the cluster to aid in efficient input and output staging and to overcome the limitations that large-scale clusters impose on local file system capacity.

In the remainder of this paper we first present an abstract model that maps collective IO concepts, previously applied in message-passing and in-memory programming environments, to the file-based MTC domain. We then review the architecture of the IBM Blue Gene/P (BG/P) system, which we use as an exemplar of large-scale clusters (and as a base for our prototype and measurements), and describe prior work on collective IO. We then describe a new collective model that addresses the challenges described above, detail its implementation, and present preliminary measurements of its performance. We conclude with an outline of our plans to extend and improve the model.

## 2 Abstract Collective IO Model for File Objects

Our abstract model, which is independent of specific cluster architectures, is based on the following elements.

1) We have applications involving multiple tasks that can run concurrently, each reading zero or more named objects, performing some computation, and writing zero or more named objects. (These objects are typically files – a detail that will become important when we talk about implementation specifics). The length of individual tasks, and of the objects read and written, are typically not known ahead of time.

2) We can distinguish between two principal input patterns: a) read-many, in which many or all tasks read the same object; b) read-few, in which the number of tasks reading a particular object is small – often only one. We assume that each object is written by just one task.

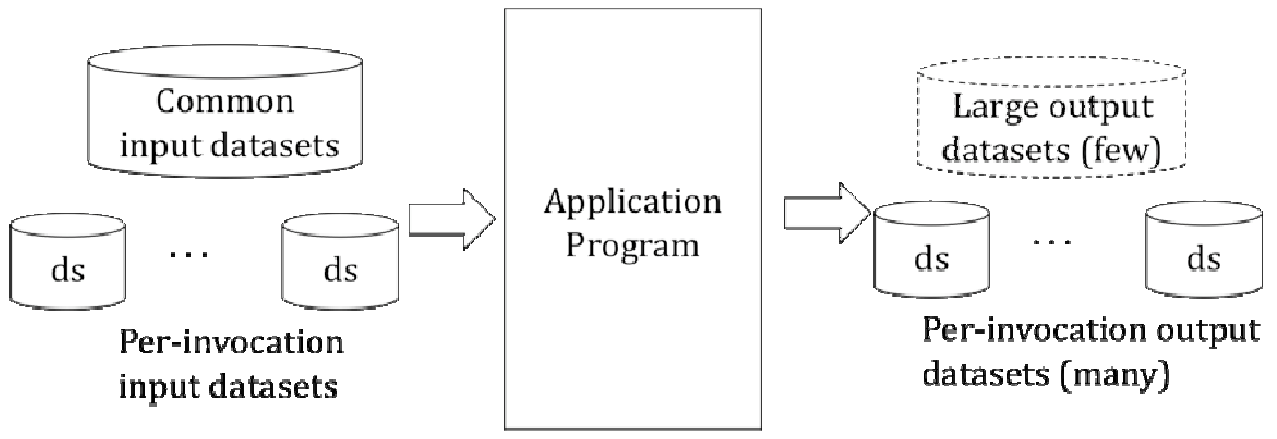

**Figure 2: Abstract application program IO profile**

Typically, we know the objects to be read by each application ahead of time, and thus assume that applications will not determine at run time which files to read. (This restriction can be relaxed for some files, which would be considered outside of, or an extension to, the model). We further assume that we know (typically, from dependency information) which objects are read-many.

3) In the simplest form of these applications, the set of objects read and the set of objects written are disjoint. In more complex forms, one task may write an object that is then read by another. In that case, we assume dataflow synchronization between the writer and the reader, meaning that the reader can only execute when the writer completes execution (as below).

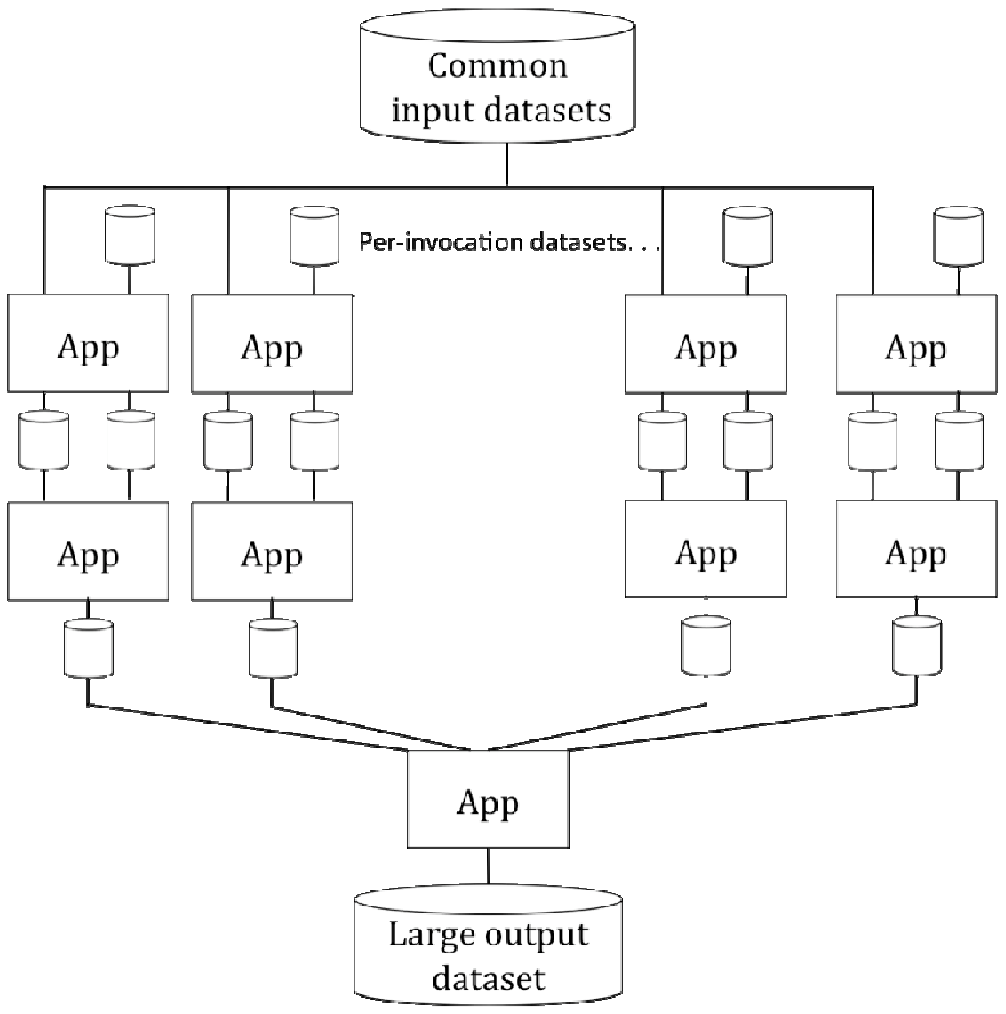

**Figure 3: Common Application Dataflow Pattern**

4) We assume a computer system architecture in which (a) all processors can access a high capacity persistent shared storage system (shared-store), albeit with modest performance, and (b) each processor has some local object storage (memory or disk) of modest capacity, but offering high performance (local-store). When many processors access the shared file system concurrently, contention leads to degraded and often unpredictable performance.

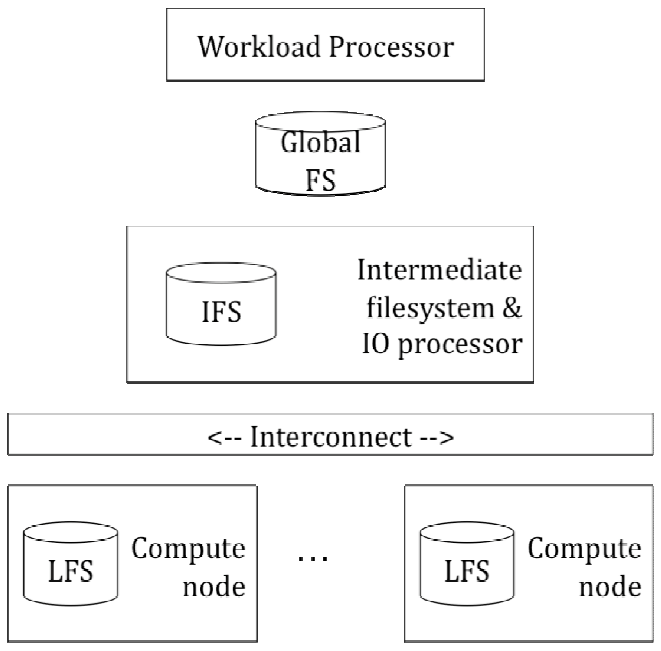

**Figure 1: Abstract Cluster**

5) An abstract cluster IO architecture is useful to define terminology. As shown in Figure 4, this model has three levels of file system: Global persistent shared file systems (GFS) are accessible from all compute nodes of a cluster, and are typically the persistent home of all data. Local file systems (LFS) are per-compute-node file systems, and are only directly accessible to tasks running on the processors of that compute node. As cluster size and density increases, the LFS may be implemented in RAM or FLASH memory, and is typically constrained in size between a few hundred megabytes and a few gigabytes. Intermediate file systems (IFS) are found, typically, only on the largest and most complex clusters, such as the IBM BGP. On the BGP, IFSs exist on the "IO node" processors (IONs); systems such as the SiCortex 5832 allow larger IFSs to be constructed by striping RAM-based LFSs. We use the acronyms GFS, LFS, and IFS throughout.

Based on this abstract model, we employ two simple collective methods to improve IO performance: (a) routines to broadcast read-many objects to many processors; and (b) two-stage IO operations to accelerate read-few and write operations, by staging objects between the many local-stores, an intermediate-store (created, for example, on a set of local-stores), and the shared-store. We implement IFSs on LFSs using MosaStore [Al-Kiswany+2007] and Chirp [Thain+2008]. Our prototype of these methods was implemented on the BG/P and is described in Section 5.

# 3 Blue Gene/P System Architecture

The 163,834-processor IBM BG/P computer at the Argonne Leadership Computing Facility [ALCF] is at the time of writing the world's largest open-science computing system [TOP500]. We view it as an exemplar of the coming wave of "petascale" systems, and we base the work described here on this system.

We present here a brief overview of the characteristics of the GPFS distributed parallel file system that serves as the GFS for the ALCF BG/P. We then describe the ZeptoOS operating system environment that we employ for MTC programming of the BG/P, as this environment is critical to enabling the MTC model to be used on this machine, and because its BG/P implementation – some of which was influenced by the work described here – has not to date been published elsewhere.

## 3.1 Characteristics of GPFS as a Global File system

GPFS – the General Parallel File system [Schmuck+2002] – is configured on the ALCF BG/P with 24 IO servers, each with 20Gb/s network connectivity, and can sustain an aggregate IO rate of ~8GB/sec.

GPFS is in general proficient at reading and writing large units, can handle vast numbers of files, and can maintain huge directories. It also excels at parallel IO operations from multiple client hosts, for which it maintains a sophisticated lock resolution protocol and heuristics. It has, however, two areas of weakness: it is relatively slow at creating new files, and can perform very poorly when multiple clients attempt to create files within the same parent directory (due to lock contention and its approach for maintaining global file system integrity in the face of metadata updates). These characteristics are typical for distributed parallel file systems which maintain local file system semantics in a distributed environment. However, they pose a challenge to MTC workloads, which can, if not carefully planned to avoid GFS weaknesses, perform exceedingly poorly.

## 3.2 BG/P OS and IO Architecture to support MTC

The ZeptoOS project [ZeptoOS] provides an open-source alternative to the proprietary software stacks available on contemporary massively parallel architectures. Its aim is to make petascale architectures more productive for the scientific user community, to enhance community collaboration and to enable computer science research on these architectures. ZeptoOS uses the Linux kernel to create an alternative, fully open software stack on large-scale parallel systems.

The project currently focuses on the IBM BG/P architecture. These machines normally run a limited microkernel on the compute nodes. While the default compute node kernel is highly scalable, it lacks many capabilities that MTC jobs expect, such as the ability to execute sub-processes or run shell scripts. ZeptoOS replaces that kernel with a Linux-based ZeptoOS compute node kernel, which lifts those limitations.

The default IBM BG/P microkernel forwards all file and socket IO calls to the IO nodes, which run Linux. IO nodes run a daemon that receives IO requests from the compute nodes and replays them against the Linux kernel. IO nodes also run file system clients for remote file systems such as NFS, GPFS, or PVFS, which handle the actual file IO.

ZeptoOS also uses a similar, but more general, forwarding architecture for IO requests. ZOID, the ZeptoOS IO Daemon [Iskra+2008], is a replacement IO daemon running on the IO nodes, used to communicate with the compute nodes when they are running Linux. ZOID provides a generic, high-performance function-forwarding infrastructure for compute nodes. This infrastructure is extensible through the use of plug-ins: users can define their own API and have data efficiently forwarded between the applications running on the compute nodes and the implementation code running on IO nodes. Generic plug-ins for POSIX file and socket IO are available which standard applications can take advantage of. ZOID also performs job management and IP packet forwarding between IO nodes and compute nodes (allowing users to, e.g., perform interactive debugging sessions on the compute nodes over telnet).

Figure 5 and Figure 6 present in more detail the hardware and software components of the ZeptoOS environment on the BG/P. The ratio of compute nodes to IO nodes for a given BG/P installation can vary from 16:1 to 128:1 depending on the machine configuration; the ratio on the Argonne machine is fixed at 64:1. Compute nodes communicate with the IO nodes over a custom "collective" (also known as "tree") network, with a bandwidth of 6.8 Gb/s (850 MB/s). Once protocol overheads are considered, the maximum throughput that ZOID can achieve over this network is around 760 MB/s. However, such throughput is only achievable when processes on the compute nodes communicate with ZOID directly. A modified GNU libc library that enables this direct communication is in progress but is currently incomplete.

A solution available to processes on the compute nodes through standard kernel interfaces would be far more desirable. Since our communication stack is in user space, we need mechanisms to forward data between the user and kernel space. The Linux kernel does offer easy to use interfaces for such purposes, in the form of FUSE and TUN. FUSE [FUSE] is a pseudo-file system that performs callbacks from the kernel VFS layer to a user-space daemon, which provides the implementation of file IO operations. TUN [TUN] simulates a network-layer device, allowing one to forward IP packets between a user-space process and the kernel's TCP/IP stack.

The problem is that neither of these solutions is particularly fast. Their designs (particularly that of FUSE) are simple and focused on flexibility, not high performance.

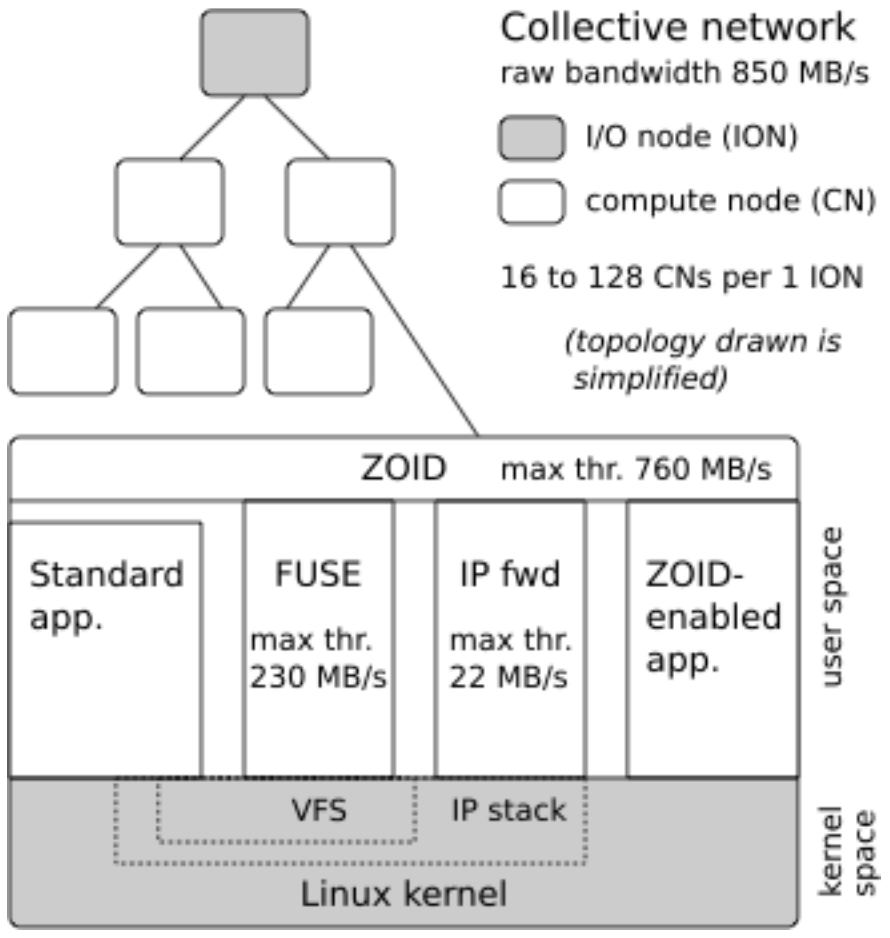

**Figure 5: ZOID and ZeptoOS**

The overheads they introduce are considerable. FUSE can read data in chunks of 128 KB, but writes are performed in chunks no larger than a single memory page. With a page size of 64 KB on the compute nodes we get at most 230 MB/s on input and 180 MB/s on output. (These are raw transfer speeds; if we include file system overhead, then even in the case of local RAM disk on the IO nodes, the read speed is reduced to 180 MB/s and the write speed to 130 MB/s).

The situation with TUN is even worse, because the data is transferred in individual IP packets of no more than 1500 bytes. As a result, we only achieve ~180 Mb/s (22 MB/s) between compute nodes and IO nodes. IP communication works between compute nodes as well, but for simplicity this is implemented in ZeptoOS by sending the packets to the IO node and letting it forward the data to the intended destination. Consequently, as the number of communicating compute node processes increases, the fraction of throughput available to each goes down.

The collective network is not the only one available on BG/P: the primary network for point-to-point communication *between* compute nodes is the 3-D torus. Every compute node has torus links to six neighbors, each with a bandwidth of 3.4 Gb/s (425 MB/s). Until recently, the torus network was not accessible when running under ZeptoOS, because the torus network's DMA engine lacks scatter/gather capability and thus requires large, continuous areas of physical memory, normally unavailable under Linux.

To enable use of the torus network under ZeptoOS, we modified the Linux kernel to reserve a considerable "flat" segment of memory at boot time. A process wishing to communicate over the torus is mapped into this memory region, so that the DMA engine can operate on its memory buffers. While this capability is still under development, we have implemented IP forwarding over MPI (which uses the torus), again using the TUN device. We measured peak torus point-to-point throughput of around 1.15 Gb/s (140 MB/s). This throughput is an order of magnitude higher than over the collective network, for several reasons, the most significant being that we have increased the maximum transmission unit (MTU) of the TUN network device to 65535 bytes (the maximum value allowed with IPv4). While we would have liked to do the same with the TUN device operating over the collective network, the older version of the Linux kernel used on the IO nodes does not allow an increase in the MTU of the TUN device. We are currently prevented from upgrading that kernel version because the GPFS kernel module depends on it.

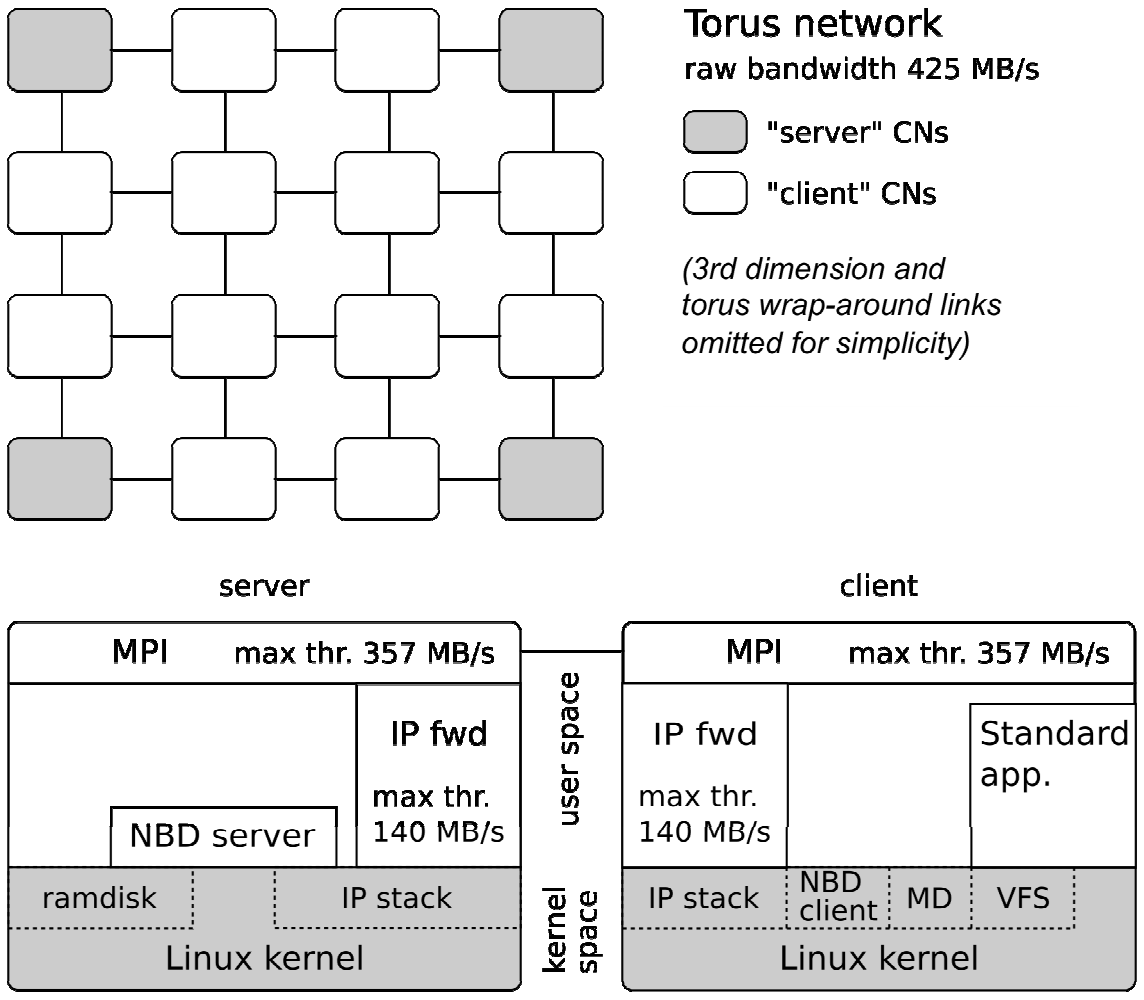

**Figure 6: ZOID/ZeptoOS and BG/P Torus Network**

## 4 Prior work

There has been much research on collective operations in the context of the message passing programming paradigm. These operations allow a group of processes to perform a common, pre-defined operation "collectively" on a set of data. For example, the MPI standard [MPI] offers a large number of such operations, from a basic broadcast (delivering an identical copy of data from one source to many destinations), through scatter (delivering a different part of input data from one source to each destination) and its opposite, gather (assembling the result at one destination from its parts available on multiple sources), to reduction operations (like gather, but instead of assembling, the parts of the result are combined). These operations are considered so crucial for the performance of message passing programs that the BG/P provides the separate collective tree network to perform them efficiently in hardware [BGP].

Similarly, *collective IO* is not a new concept in parallel computing. It is employed, e.g., by ROMIO [Thakur+1999], the most popular MPI-IO implementation, in its generalized two-phase IO implementation. When compute tasks want to perform IO, they first exchange information about their intentions, in an attempt to coalesce many small requests into fewer larger ones (an assumption being that the processes access the same file). When reading, in the first phase the processes issue large read requests, and in the second phase, they exchange parts of their read buffers with one another, using efficient MPI communication primitives so that each process ends up with the data it was interested it. For writing, the two phases are reversed.

MPI collective communication and IO operations require applications to be at least loosely synchronous, in that progress must be made in globally synchronized phases, and that all processes participate in a collective operation. These conditions restrict the use of standard collective operations in loosely coupled, uncoordinated scenarios, limiting them to initialization time (before any individual tasks start running), and possibly termination time (once all individual tasks have completed).

Until recently, such uncoordinated jobs were primarily run on moderate scale clusters or on distributed ("grid") resources. Most clusters were not large enough to encounter IO contention problems such as those described here. Furthermore, cluster nodes generally have considerable local disks suitable for storing large input and output data. The primary problem on such systems has thus been mainly to efficiently stage data and schedule jobs so that they can best benefit from the staged data [Khanna+2006; Khanna+2007].

File IO is a more significant problem with distributed resources. Condor provides a remote IO library that forwards system calls to a shadow process running on the "home" machine where the files actually reside. Global Access to Secondary Storage (GASS [Bester+1999]) available in Globus takes a different approach, transparently providing a temporary replica cache for input and output files. Our collective IO goes beyond these approaches to intelligently utilize local filesystems, and to provide intermediate file systems, broadcasting of input files, and batching of output files. Unlike Condor remote IO, our approach does not require relinking. Our approach makes it practical for tens to hundreds of thousands of processor cores now (and in a few years, a million cores) to perform concurrent, asynchronous IO operations. These numbers are easily an order of magnitude greater than what has been addressed in any previous implementation.

## 5 Design and Implementation

The requirements described to this point translate into a straightforward design for handling collective IO, which consists of three main components: 1) one or more *intermediate file systems* (IFSs) enabling data to be placed and cached closer to the computation (from an access-latency and bandwidth perspective) while overcoming the size limitation of the typical RAM-based local file systems that are prevalent in petascale-precursor systems; 2) a *data distributor*, which replicates sufficiently large common input datasets to intermediate file systems; and 3) a *data collector* mechanism, which collects output datasets on IFSs and efficiently writes the collected data to large archive files on the GFS.

Our implementation of this design, which we have prototyped for performance evaluation, uses simple scripts to coordinate "off the shelf" data management components. All of our prototypes and measurements to date have been done on the Argonne BG/P systems (Surveyor, 4096 processors, and Intrepid, 163,840 processors). Not all of the design aspects described below exist yet in the prototype. These are indicated in the description. We executed all of our compute tasks under the Falkon lightweight task scheduler [Raicu+2007; Raicu+2008] running under ZeptoOS [ZeptoOS2008].

The structure of the system is shown in overview in Figure 7, and in more detail in Figure 9, which depicts the flow of input and output data in our BG/P-based prototype. Within the BG/P testbed, the RAM-based file system of the local node, which contains about 1GB of free space, is used as the LFS. For input staging, the LFS of one or more compute nodes is set aside as a "file server" and is dedicated as an IFS for a set of compute nodes.

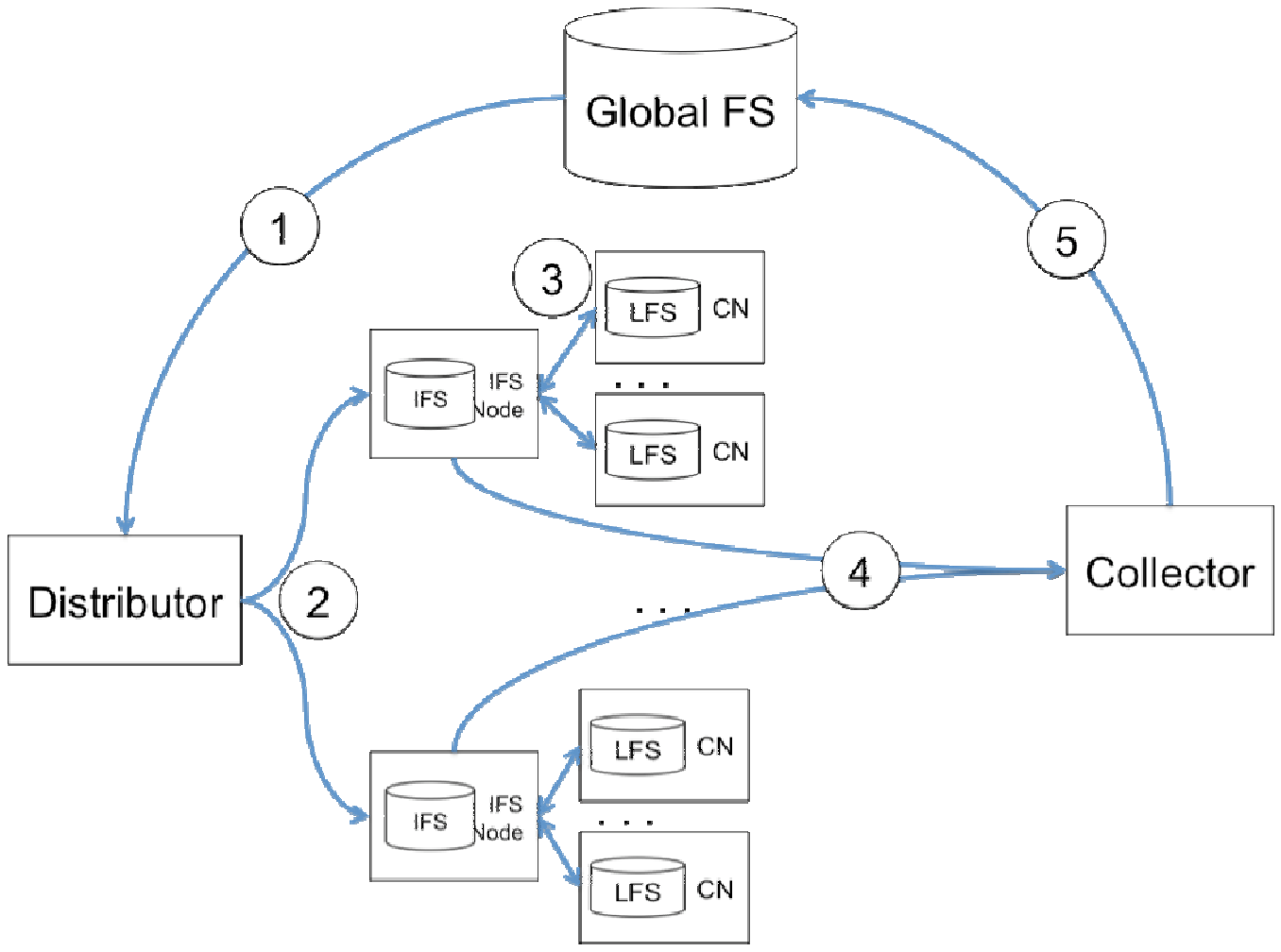

**Figure 7: Logical Distributor/Collector Design**

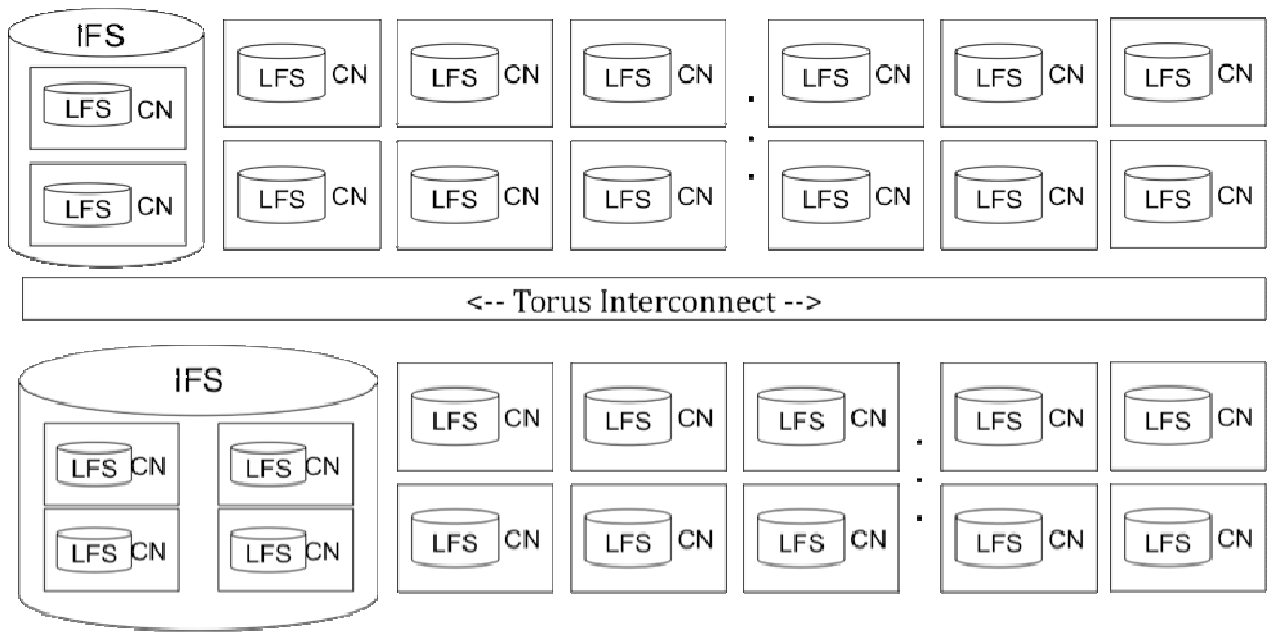

**Figure 8: Allocation and mapping of compute nodes to IFS servers: 2:64 ratio (top) and 4:64 ratio (bottom)**

We create large IFSs from fast LFSs by striping IFS contents over several LFS file systems, using the MosaStore file IO service [Al-Kiswany+2007]. Compute nodes access the IFS using the BG/P torus network [BGP]. The creation of the IFS and the partitioning of compute nodes between IFS functions and computing can be done on a per-workload basis, and can vary from workload to workload. In the same manner that compute node and IO node operating systems are booted when a BG/P job is started, the creation of the IFSs and the CN-to-IFS mapping can be performed as a per-workload setup task performed when compute nodes are provisioned by Falkon [Raicu+2007; Raicu+2008]. This enables the CN-to-IFS ratio to be tailored to the disk space and bandwidth needs of the workflow (Figure 8).

### 5.1 Input Distribution

The input distributor stages common input data efficiently to LFS or IFS. This mechanism is used to cache files that will be frequently re-read, or that will be read in inefficient buffer lengths, closer to the compute nodes. The key to this operation is to use broadcast or multicast methods, where available, to move common data from global to local or intermediate file systems. For accessing input data, we stage input datasets as follows:

- Small input datasets are staged from GFS to the LFS of the compute nodes which read them.
- Datasets read by only one task but that are too large to be staged to an LFS are staged to an IFS of sufficient size.
- All large datasets that are read by multiple tasks are replicated to all IFSs that serve the set of compute nodes involved in a computation.

In our prototype implementation, data is replicated from GFS to multiple IFSs by the Chirp *replicate* command [Thain+2008]. (Steps 1 and 2 in Figure 7.) We employ two functions: the first identifies if a given compute node is a data-serving or application-executing node. The second maps executor compute nodes to its IFS data server. The decision of whether to place an input file on LFS or IFS is made explicitly (i.e., hard-coded in our prototype). Each IFS is mounted on all associated compute nodes, and accessed via FUSE.

### 5.2 Output Collection

The output collector gathers (small) output data files from multiple processors and aggregates them into efficient units for transfer to GFS. In this way, we reduce greatly the number of files created on the GFS (which reduces the number of costly file creation operations) and also increase the size of those files (which permits data to be written to GFS in larger, more efficient block sizes and write buffer lengths). The use of the output collector also enables data to be cached on LFS or IFS for later analysis or reprocessing.

Our goal is that files which can fit on the LFS can be written there by the application program, while larger output files can be written directly to IFS, and output files too large to fit on the LFS or IFS are written directly to GFS. (This differentiation is not implemented in the prototype). In this way, we can optimize the performance of output operations such as file and directory creation and small write operations.

The collector operates as follows. When application programs complete, any output data on the LFS is copied to an IFS (Figure 7, Step 3). When the copy is complete, the data is atomically moved to a staging directory, where the following algorithm (Step 4) is used:

```
while workload is running
  if time since last write > maxDelay
     or data buffered > maxData
     or free space on IFS < minFreeSpace
  then write archive to GFS from staging dir
```

One consequence of this design is that short tasks can complete quicker, without having each task remain on a compute node waiting for its data to be written to GFS, as the staging of data from IFS to GFS is handled asynchronously by the collector, as shown in Figure 10. In our prototype, the IO node (ION) file system serves as the IFS, and data moving relies on POSIX atomicity semantics for data integrity. Files are moved from LFS to IFS via tar, and are then transferred to GFS using dd with a large efficient blocksize.

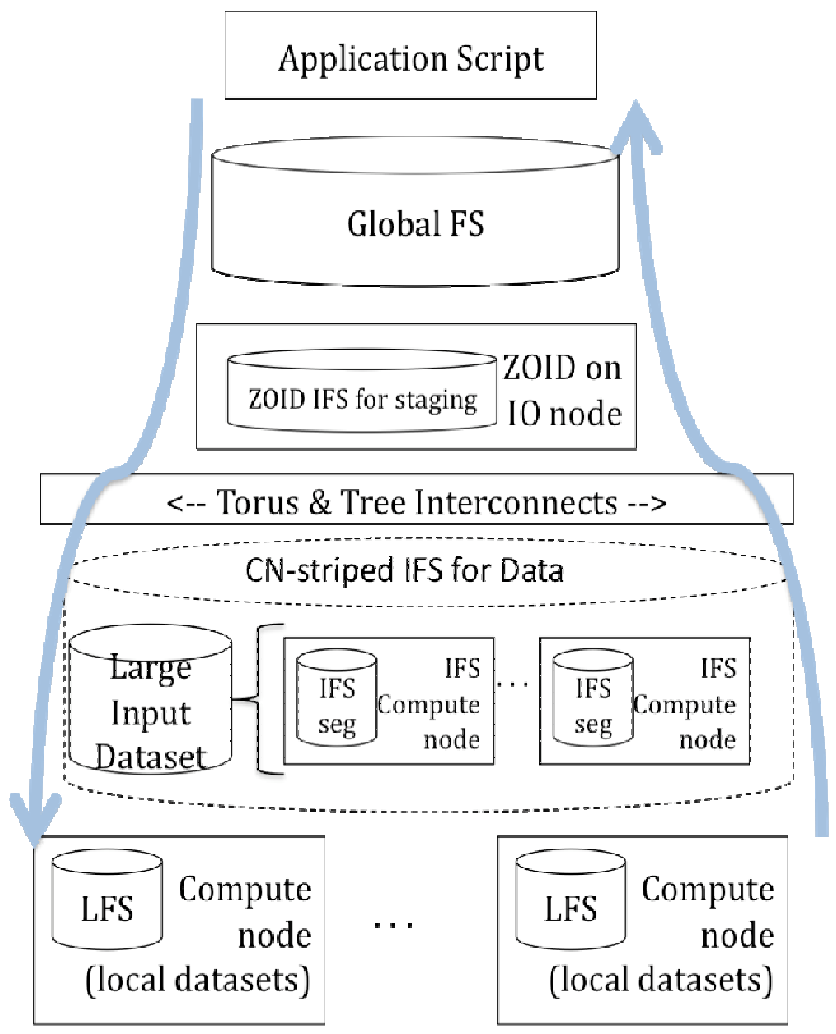

**Figure 9: Data flow on BG/P**

In our prototype, the LFS and IFS file systems are both RAM-based, and behave somewhat like an in-memory message exchange system, in which messages are moved by read() and write() from one namespace (file server) to another. While these "messages" may be more expensive than MPI messages (the difference remains to be measured), this approach lets users integrate existing application programs into larger application workflows without requiring disk IO.

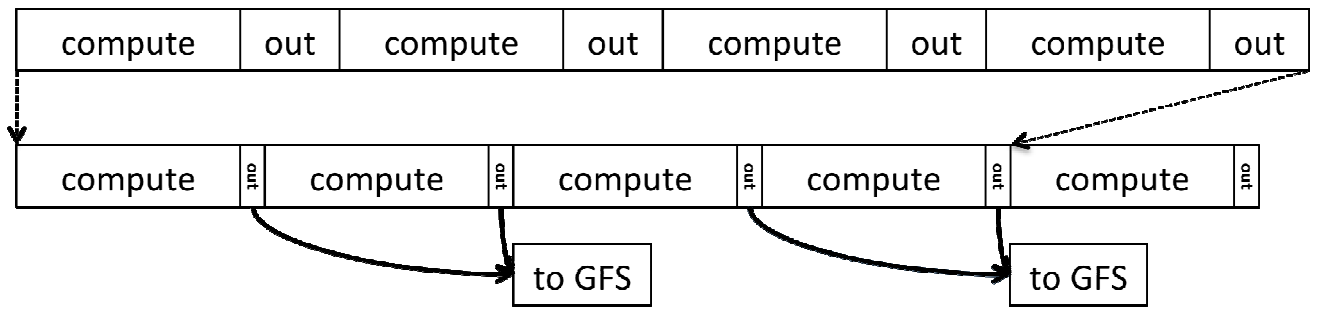

**Figure 10: Output staging: synchronous, top, without collector; asynchronous, bottom with collector.**

### 5.3 Downstream data processing

The fact that data managed by the output collector on LFSs or IFSs can be retained for subsequent processing makes it possible to re-process the output data of one stage of a workflow far more efficiently than if the data had to be retrieved from GFS. When previously written output does need to be retrieved from GFS, the ability to access files in parallel from a randomly accessible archive (as described below) further improves performance. And intermediate output data that doesn't need to be retained persistently can be left on LFS or IFS storage without moving it to GFS at all.

To facilitate multi-stage workflows, in which the output of one stage of a parallel computation is consumed by the next, we incorporate two capabilities in our design: 1) the use of an archive format for collective output that can be efficiently re-processed in parallel, and 2) the ability to cache intermediate results on LFS and/or IFS file systems.

We base our output collector design on the use of a relatively new archive utility xar [XAR], which unlike traditional tar (and similar) archive formats includes an updateable XML directory containing the byte offset of each archive member. This directory enables files to be extracted via random access, and hence xar (unlike tar) archives can be processed efficiently in parallel in later stages or a workflow. In the future, it is likely that we can implement parallel IO to an xar archive from multiple compute nodes, thus enhancing write performance potential even further. To enable testing of such re-processing of derived data from LFS, we employ a prototype of a new primitive collective execution operation "run task x on all compute nodes" which enables all previous outputs on LFS to be processed. Our prototype does not yet use xar, but rather tar, which has a similar interface.

# 6 Performance Evaluation

We present measurements from the Argonne ALCF BG/P, running under ZeptoOS and Falkon. We have evaluated various features on up to 98,304 (out of 163,840) processors. Dedicated test time on the entire facility is rare, so all tests below were done with the background noise of activity from other jobs running on other processors. Nonetheless, the trends indicated are fairly clear, and we expect that they will be verifiable in future tests in a controlled, dedicated environment. We have made measurements in both areas of the proposed collective IO primitives (denoted as CIO throughout this section), such as *input* data distribution, and *output* data collection. We also applied the collective IO primitives to a molecular dynamics docking application at up to 96K processors.

## 6.1 Input Data Distribution

Our first set of results investigated how effectively compute nodes can read data from the IFSs (over the torus network), examining various data volumes and various IFS/LFS ratios. We used the lightweight Chirp file system [Thain+2008] and the Fuse interface to read files from IFS to LFS. Figure 11 shows higher aggregate performance with larger files, and with higher ratios, with the best IFS performance reaching 162 MB/s for 100 MB files and a 256:1 ratio. However, as the bandwidth is split between 256 clients, the per-node throughput is only 0.6 MB/s. Computing the per-node throughput for the 64:1 ratio yields 2.3 MB/s, a significant increase. Thus, we conclude that a 64:1 ratio is good when trying to maximize the bandwidth per node. Larger ratios reduce the number of IFSs that need to be managed; however, there are practical limits that prohibit these ratios from being extremely large. In the case of a 512:1 ratio and 100 MB files, our benchmarks failed due to memory exhaustion when 512 compute nodes simultaneously connected to 1 compute node to transfer the 100 MB file. This needs further analysis.

Our next set of experiments used the lightweight MosaStore file system [Al-Kiswany+2007] to explore how effectively we can stripe LFSs to form a larger IFS. Our preliminary results in Figure 12 show that as we increase the degree of striping we get significant increases in aggregate throughput, up from 158 MB/s to 831 MB/s.

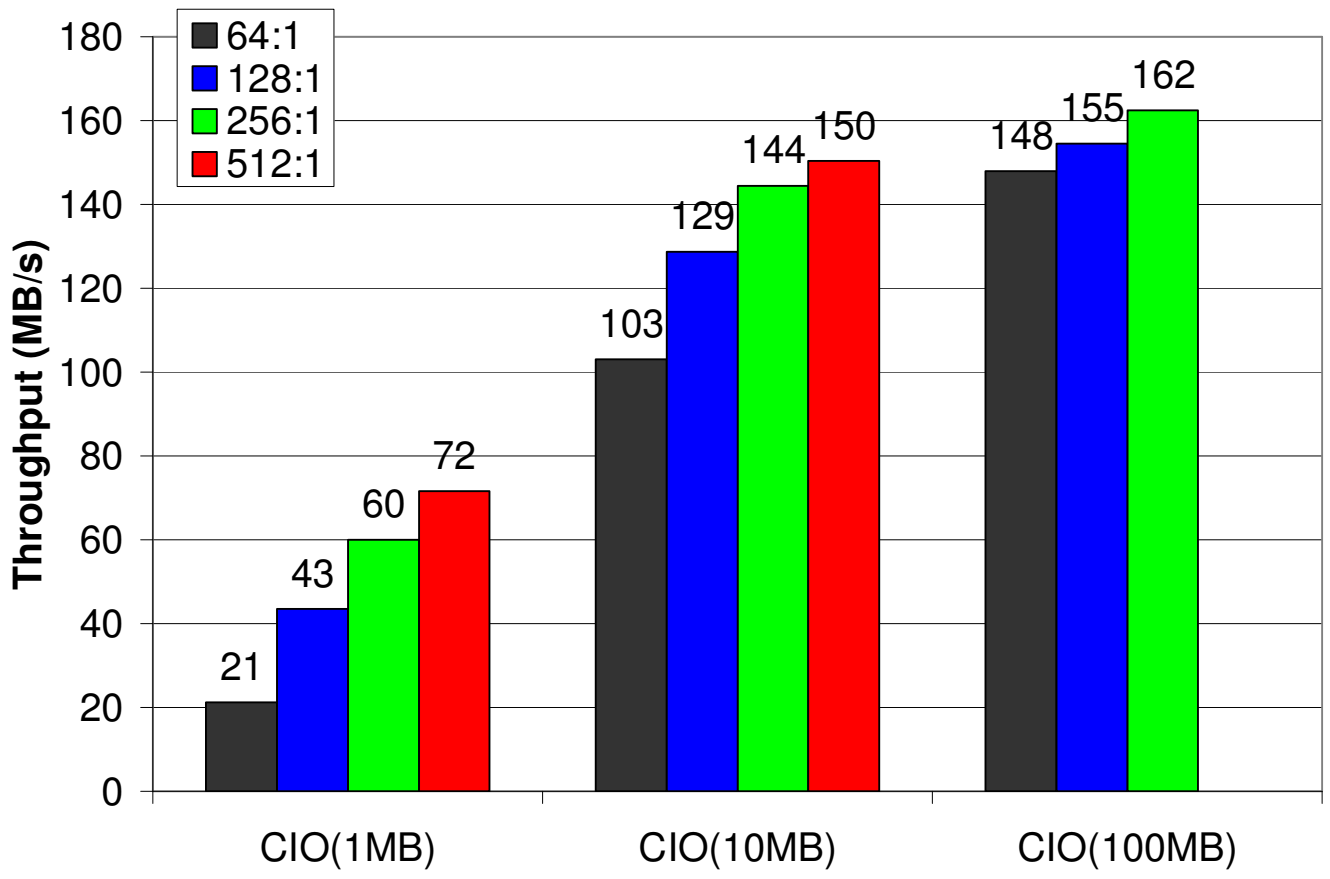

**Figure 11: Read performance while varying the ratio of LFS to IFS from 64:1 to 512:1 using the Torus network.**

The best performing configuration was 32 compute nodes aggregating their 2GB-per-node LFSs into a 64 GB IFS. This aggregation not only increases performance, but also allows compute nodes to keep their IO relatively local when working with large files that do not fit in a single compute node 2GB RAM-based LSF.

Our final experiment for the input data distribution section focused on how quickly we can distribute data from GFS to a set of IFSs, or potentially to LFSs. As in our previous experiment, we use Chirp (see Figure 13). Chirp has a native operation that allows a file (or set of files) to be distributed to a set of nodes over a spanning tree of copy operations. The spanning tree has the benefit of requiring fewer data transfers: log($n$) instead of $n$, where n is the number of nodes.

In the case of a naïve data distribution in which compute nodes read data directly from GFS (GPFS in our case as noted in the figure), computing the aggregate throughput is straightforward: *throughput = nodes*dataSize/workloadTime.* For the spanning tree distribution, computing the actual throughput is problematic since the number of transfers is lower than in the naïve method. To make the comparison fair, we compute throughput for the spanning tree distribution with the same formula as for the naïve data distribution, although the actual network traffic would have been significantly less. We believe this is the correct way to compare the two approaches, as it emphasizes the time to complete the workload. On up to 4K processors, GPFS achieves 2.4 GB/s at the largest scale (2.4 MB/s per node). This is the peak rated performance for the file system we tested *(/home*). However, the spanning tree approach achieves an equivalent of 12.5 GB/s on 4K processors. We plan to explore the performance of the spanning tree distribution at larger scales to find the torus network saturation point. We expect to achieve at least one order of magnitude better performance (for distributing a set of files to many compute nodes) at large scales when using the spanning tree approach as opposed to the naïve approach which reads each file from GPFS directly.

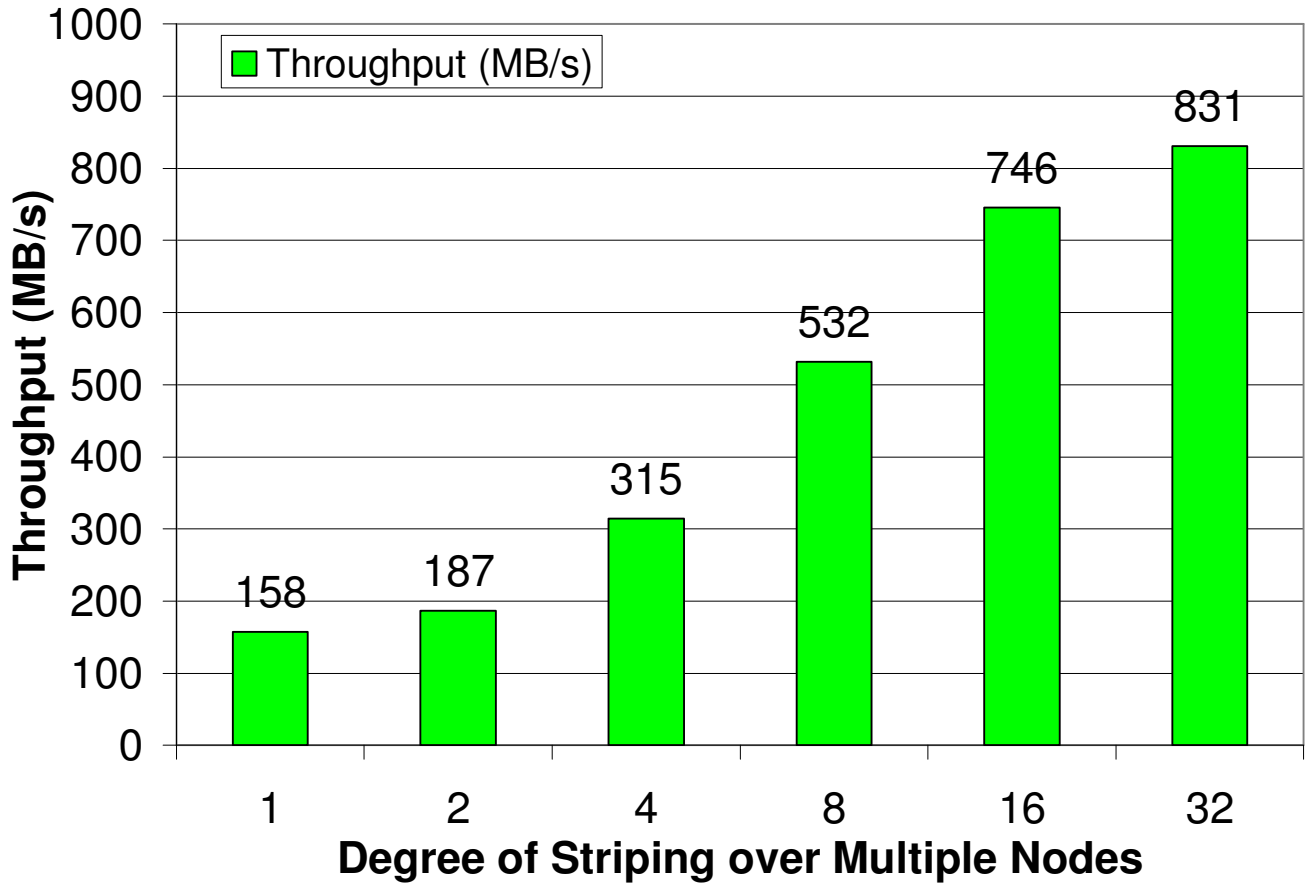

**Figure 12: Read performance, varying the degree of striping of data across multiple nodes from 1 to 32 using the torus network**

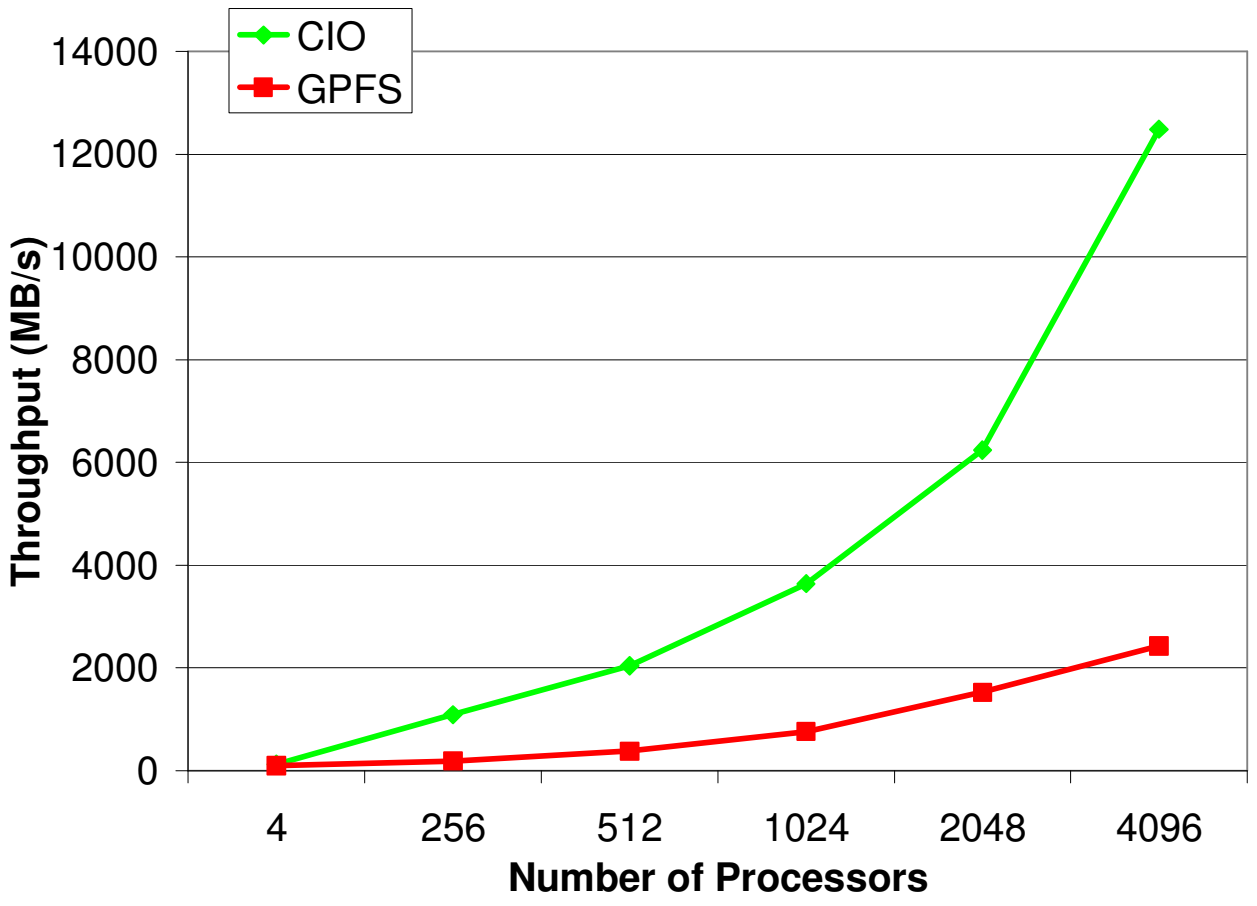

**Figure 13: CIO distribution via spanning tree over Torus network vs. GPFS over Ethernet & Tree networks**

## 6.2 Output Data Collection

Our second goal for the collective IO primitives was to support the aggregation and transfer of many files from multiple LFSs or IFSs to the GFS. When writing from many compute nodes directly to GPFS (the GFS on the BG/P), care must be taken to avoid locking contention on metadata. One way to avoid this problem is to ensure that each compute node writes files to a unique directory. It is desirable to have as few clients as possible writing to GFS concurrently to limit any locking contention, and to allow the largest buffer sizes and aggregation and potentially small files into larger ones. It is also desirable to make write operations as asynchronous as possible to allow the overlap of computing and data transfer from the compute node. To achieve all these desirable features, we have implemented an output data collector (CIO, which we previously discussed) that resides on an IFS and acts as an intermediate buffer space for output generated on compute nodes. We use a ratio of 64:1 IFS to LFS, which significantly reduces the number of clients that write to GFS.

Our measurements (see Figure 14 and Figure 15) show that the CIO collector strategy yields close to the ideal efficiency when compared to compute tasks of the same length with no IO. For example, in Figure 14 we show the efficiency achieved with short tasks (4 seconds) that produce output files with sizes ranging from 1KB to 1MB. We see that CIO (the dotted lines) is able to achieve > 90% efficiency in most cases, and almost 80% in the worst case with larger files. In contrast, the same workload achieved only 10% to < 50% efficiency when using GPFS. We also observed an anomaly: a slight efficiency increase at the largest scale of 32K processors. One possible cause of this is that we reached the limit of Falkon dispatch throughput.

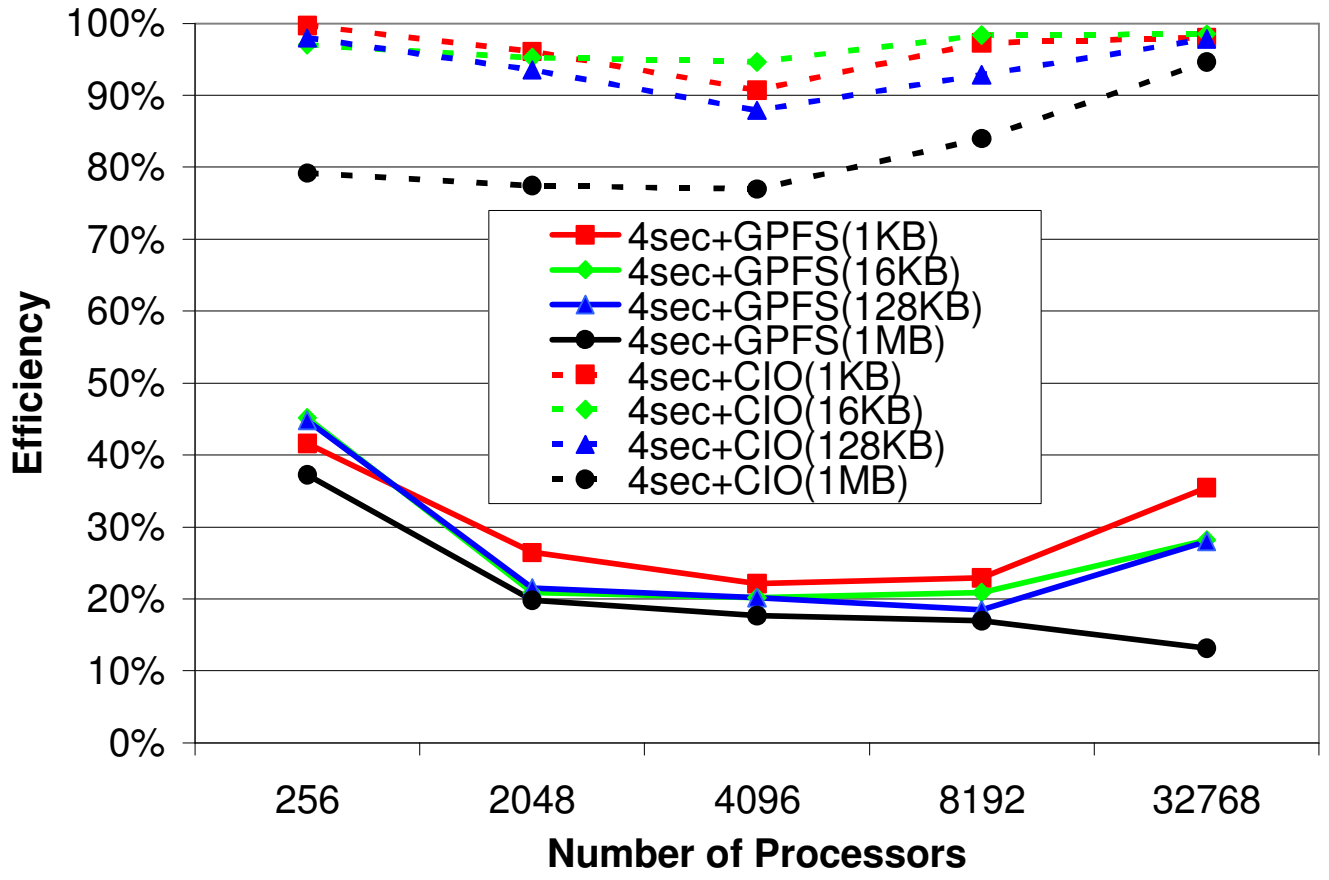

**Figure 14: CIO vs. GFS efficiency for 4 second tasks, varying data size (1KB to 1MB) on 256 to 32K processors**

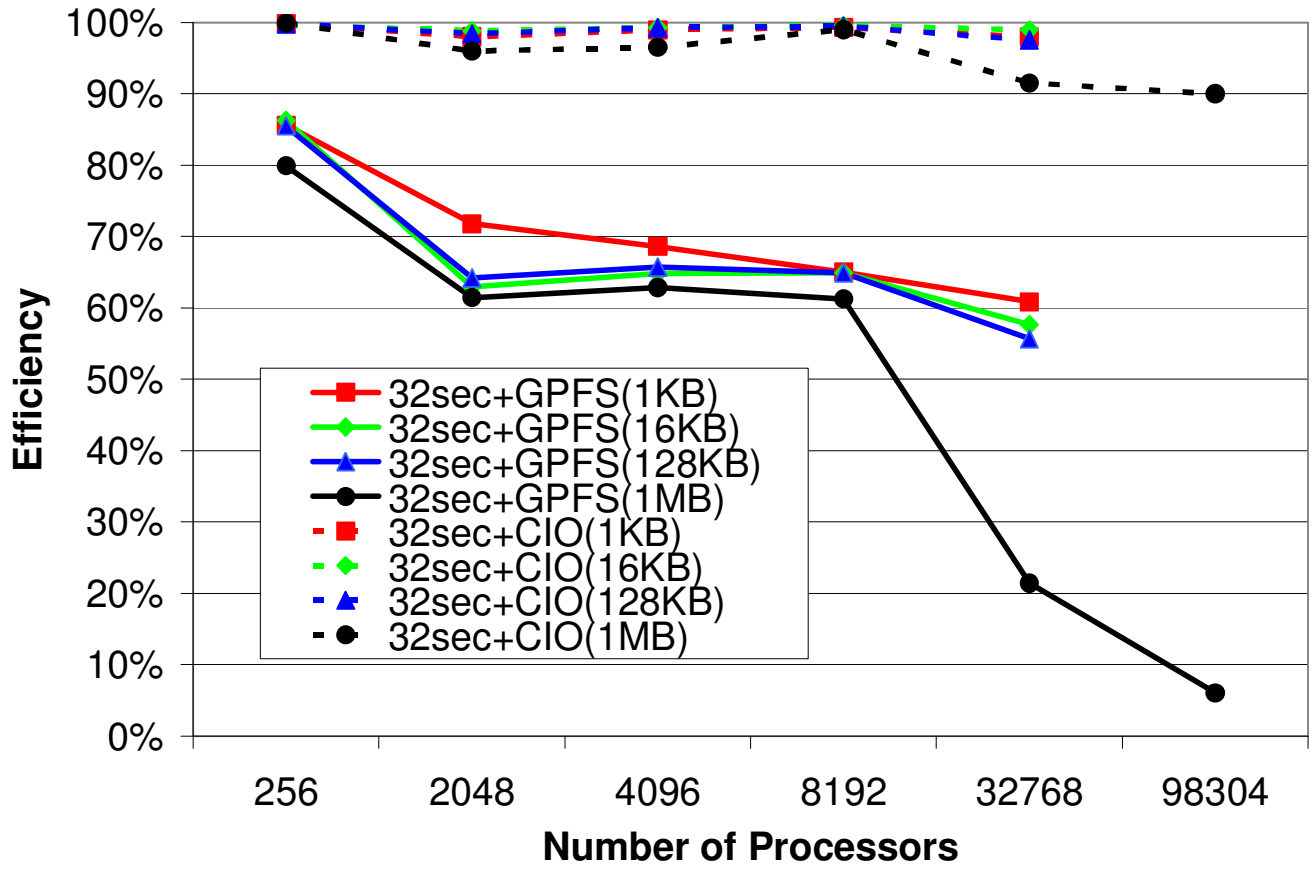

**Figure 15: CIO vs GPFS efficiency for 32 second tasks, varying data size (1KB to 1MB) for 256 to 96K processors.**

Figure 15 is similar to Figure 14, but uses 32 second tasks. We see a similar pattern, in which CIO achieves 90% efficiency, while GPFS achieves almost 90% efficiency with 256 processors but less than 10% on 96K processors.

We also extract from these experiments the achieved aggregate throughput (shown in Figure 16). We limit this plot to the 1 MB case for readability. Notice the extremely poor GPFS write performance as the number of processors increases, peaking at only 250 MB/s. The CIO throughput is almost an order of magnitude higher, peaking at 2100 MB/s, and is within a few percent of the ideal case (tasks with the same duration, but with only local IO to RAM-based LFS, labeled 4sec+RAM and 32sec+RAM).

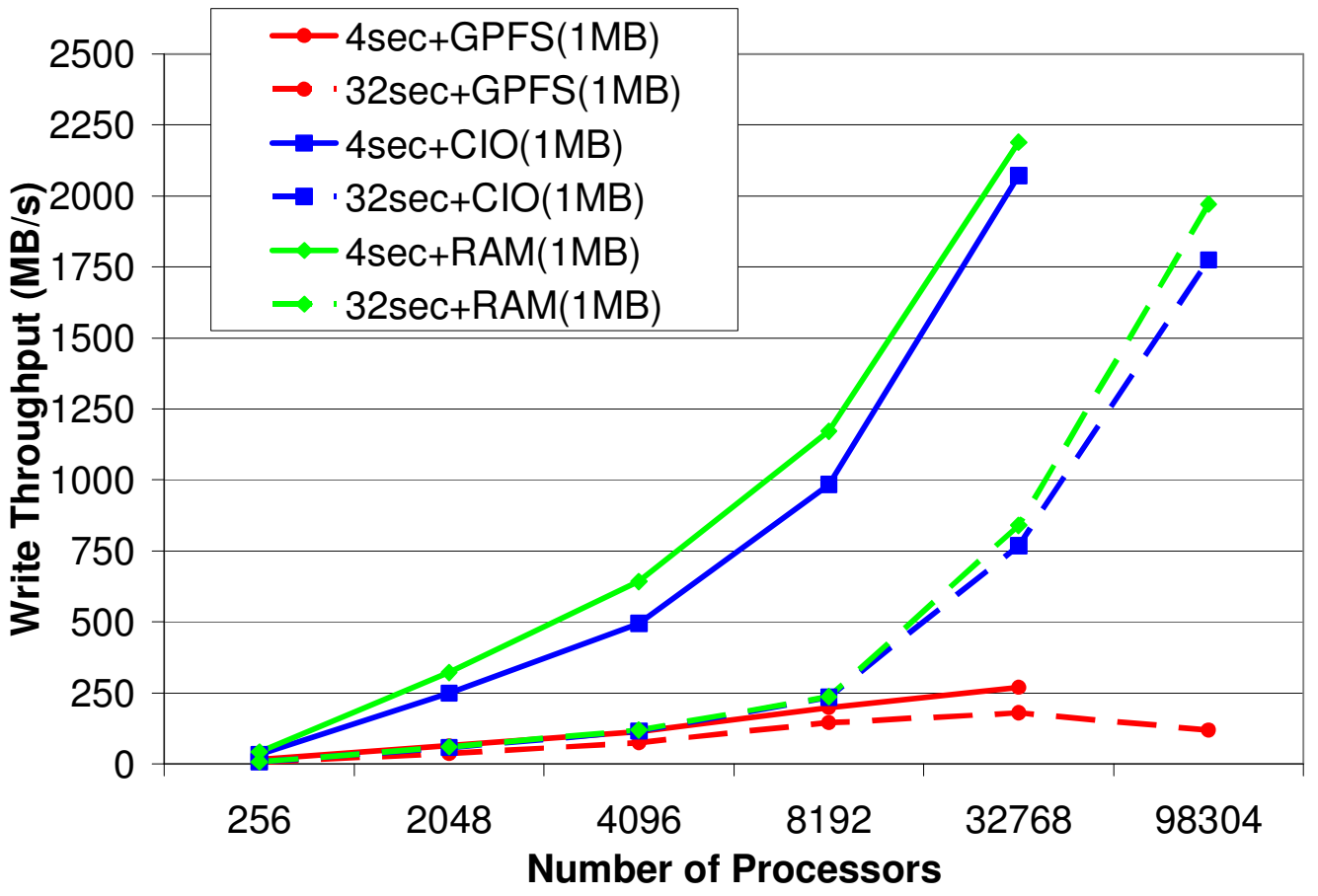

**Figure 16: CIO collection write performance compared to GPFS write performance on up to 96K processors**

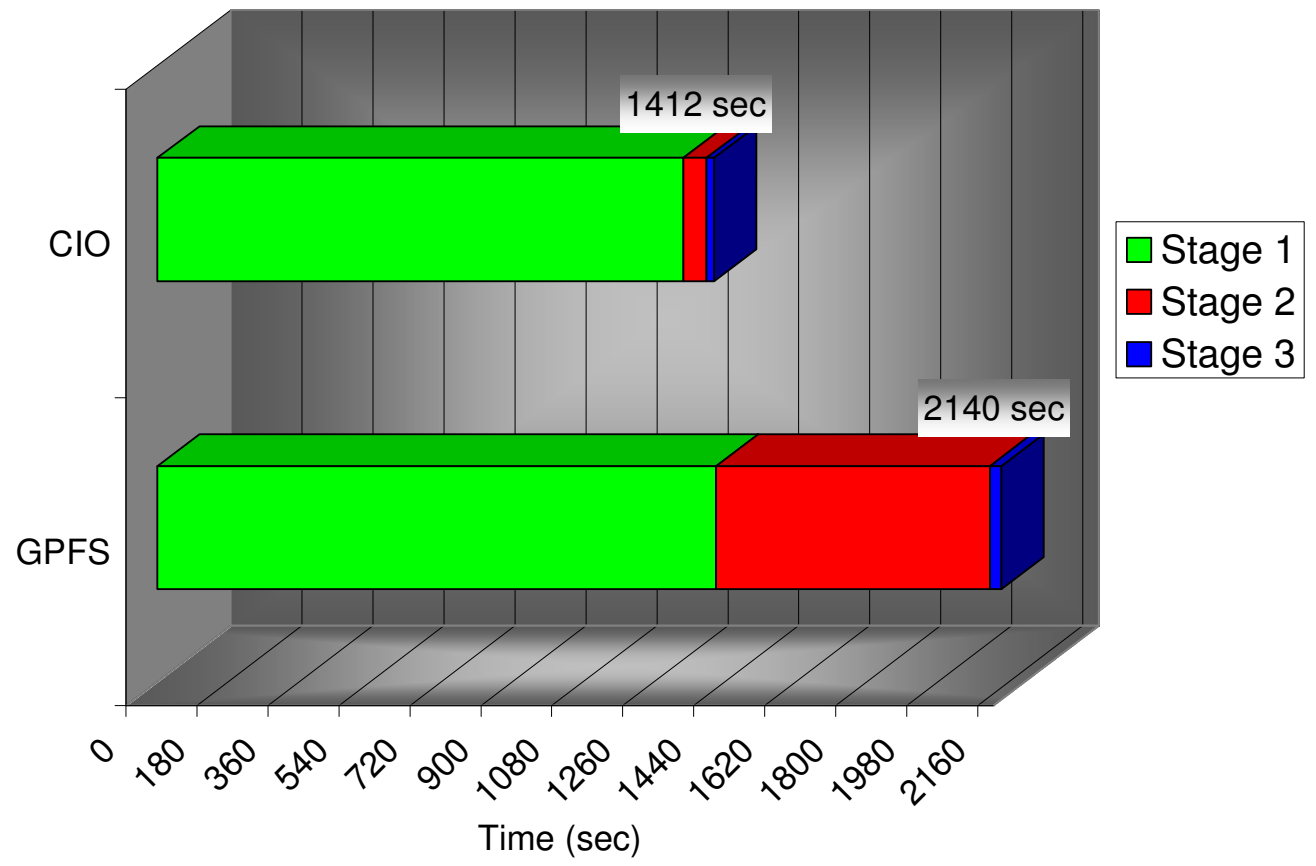

**Figure 17: DOCK6 application summary with 15K tasks on 8K processor comparing CIO with GPFS**

### 6.3 Application Evaluation

We have shown significant performance and scalability improvements for synthetic data-intensive workloads. To determine how these improvements translate into real application performance, we evaluated the utility of collective IO on a molecular dynamics workflow which screens candidate drug compounds against metabolic protein targets using the DOCK6 application [DOCK] to simulate the "docking" of small molecules to the "active sites" of large macromolecules. A compound that interacts strongly with a receptor, such as a protein molecule, associated with a disease, may inhibit its function and thus act as a beneficial drug. In this application run, a database of 15,351 compounds was screened against nine proteins that perform key enzymatic functions in the metabolism of bacteria and humans.

The molecular dynamics docking workflow has 3 stages: 1) read input, compute the docking, and write output; 2) summarize, sort, and select results; and 3) archive results. In out tests, the DOCK6 invocations averaged 10KB of output every 550 seconds.

In the simple case where we use GFS, the input data of stage 1 is read from GFS to LFS, the application reads from LFS and writes its output to LFS, and finally the output is synchronously copied back to GFS. Stage 1 is parallelized to process each DOCK invocation on a separate processor core. Both stage 2 and stage 3 were originally a single process application that would run on a login node and access input data directly from GFS. In the case of using CIO, the stages are a bit different: stage 1 writes the output data from LFS to IFS asynchronously; stage 2 is parallelized across all processors and works on IFS; stage 3 copies the data from IFS to GFS. Figure 17 shows the breakdown of the 3 stages, and where time was being spent, for a total of 1412 seconds for CIO and 2140 seconds for GPFS. The first stage is negligibly faster with CIO (1.06X), and the third stage is 1.5X faster, but the second stage is 11.7X faster with 694 seconds being reduced down to 59 seconds. Stage 2 summarizes, sorts and filters the results, which CIO can handle much better in a distributed fashion (as opposed to the centralized GFS solution) with data accesses localized to IFS instead of GFS.

In order to see the effects of CIO at larger scale, we also ran the DOCK6 stage 1 with 135K tasks on 96K processors. The net result was a 1.12X speedup using CIO (1772 seconds) as compared to GPFS (1981 seconds) – a negligible speedup, as we expected for this compute-bound workload.

# 7 Future Work

The prototype implementation we describe here, while in its early stages of development, has been sufficient to make a reasonable assessment of the performance and usability potential of a file-based collective IO model that can handle at least O(100K) BG/P processors. Our next major focus will be to integrate the model into the Swift parallel programming environment [Zhao+2007], so that BG/P users can benefit from this higher-level programming model without explicitly programming the collective IO operations.

We intend to investigate algorithmic questions and enhancements, such as determining the optimal ratio of IFS nodes to compute nodes for various workloads; determining when we can effectively use the compute nodes of IFS data hosts for computing in addition to file serving; automatically optimizing input data placement on LFSs vs. IFSs; determining if we can learn from the IO patterns of previous runs where best to locate a given input or output file; finding algorithms for automating output data caching in IFSs and LFSs for re-processing by subsequent workflow stages; and determining when data on IFSs/LFSs can be removed.

Lower-level implementation issues we intend to explore include the use of the tree network to enhance the performance of input broadcast, and comparing the performance and reliability benefits of MosaStore, Chirp, and native Linux approaches to IFS striping. We also intend to explore how the random access capabilities of archive formats such as xar can enable parallel reading and parallel archive creation, and what role compression should play in the output process.

We will continue to drive this work with an expanding measurement effort, on both synthetic and actual applications. We are particularly interested in measuring the behavior of applications (such as BLAST runs on large databases) that will benefit greatly from striped IFS capabilities.

# 8 Conclusion

We have identified, characterized, and started to address a critical problem for enabling the use of petascale supercomputers by a far larger community of scientific applications and users: how to enable efficient file-based IO by large numbers of independent parallel tasks, as required by many-task computing applications involved in loosely coupled parallel programming.

Our results indicate that it is possible to adapt principles of collective data operations to the world of parallel scripting linked by file interchange. While our results are preliminary, and are based on simple prototypes, they suggest that collective IO primitives, when effectively integrated into parallel scripting programming systems and languages (such as Falkon and Swift) can yield excellent performance on 100,000 processors – and likely well beyond – while greatly enhancing scientific programming productivity.

**ACKNOWLEGEMENTS**

This work was supported in part by the National Science Foundation under Grant OCI-0721939, by NASA Ames Research Center GSRP Grant Number NNA06CB89H, and by the Mathematical, Information, and Computational Sciences Division subprogram of the Office of Advanced Scientific Computing Research, Office of Science, U.S. Dept. of Energy, under Contract DE-AC02-06CH11357.

The authors would like to thank Samer Al-Kiswany of the University of British Columbia for assistance with MosaStore, Kazutomo Yoshii of Argonne National Laboratory for assistance with ZeptoOS, the Argonne Leadership Computing Facility team for their tremendous support in our use of the Intrepid BG/P, and Mike Kubal of the Computation Institute for providing and explaining the molecular docking workflow.

**REFERENCES**

**[ALCF]** Argonne Leadership Computing Facility, http://www.alcf.anl.gov

**[Al-Kiswany+2007]** S. Al-Kiswany, M. Ripeanu, S. Vazhkudai, "A Checkpoint Storage System for Desktop Grid Computing", Networked Systems Lab, U. of British Columbia, Tech Report NetSysLab-TR-2007-04, 2007.

**[Bester+1999]** J. Bester, I. Foster, C. Kesselman, J. Tedesco, and S. Tuecke, "GASS: A data movement and access service for wide area computing systems", IOPADS 99: Proceedings of the Sixth Workshop on IO in Parallel and Distributed Systems, Atlanta, GA, pp 78-88, 1999.

**[BGP]** IBM Blue Gene team, "Overview of the IBM Blue Gene/P Project". IBM Journal of Research and Development, vol. 52, no. 1/2, pp. 199-220, Jan/Mar 2008.

**[DOCK]** Overview of DOCK, http://dock.compbio.ucsf.edu/Overview_of_DOCK/index.htm

**[FUSE]** FUSE: File System in Userspace. http://fuse.sourceforge.net/

**[Iskra+2008]** K. Iskra, J. W. Romein, K. Yoshii, and P. Beckman. "ZOID: IO-forwarding infrastructure for petascale architectures". 13th ACM SIGPLAN Symposium on Principles and Practice of Parallel Programming, pp. 153-162, Salt Lake City, UT, Feb. 2008.

**[Khanna+2006]** G. Khanna, N. Vydyanathan, U. V. Catalyurek, T. M. Kurc, S. Krishnamoorthy, P. Sadayappan, J. H. Saltz, "Task Scheduling and File Replication for Data-Intensive Jobs with Batch-shared IO", Proceedings of the 15th IEEE International Symposium on High-Performance Distributed Computing (HPDC-15) pp. 241-252, June 2006.

**[Khanna+2007]** G. Khanna, U. V. Catalyurek, T. M. Kurc, P. Sadayappan, J. H. Saltz, "Scheduling File Transfers for Data-Intensive Jobs on Heterogeneous Clusters", Proceedings of Euro-Par 2007 Parallel Processing, pp. 214-223, August, 2007.

**[MPI]** Message Passing Interface Forum, "MPI-2: Extensions to the Message-Passing Interface", http://www.mpi-forum.org/docs/mpi-20-html/mpi2-report.html

**[MPI-IO]** K. Coloma, A. Ching, A. Choudhary, W. Liao R. Ross, R. Thakur, L. Ward, "A New Flexible MPI Collective IO Implementation", International Conference on Cluster Computing, 2006.

**[NBD]** Network Block Device. http://nbd.sourceforge.net/

**[Ousterhout1998]** J. Ousterhout, "Scripting: Higher-level programming for the 21st century", IEEE Computer Mar. 1998.

**[Raicu+2007]** I. Raicu, Y. Zhao, C. Dumitrescu, I. Foster, M. Wilde. "Falkon: a Fast and Light-weight tasK executiON framework", IEEE/ACM Supercomputing 2007.

**[Raicu+2008]** I. Raicu, Z. Zhang, M. Wilde, I. Foster, P. Beckman, K. Iskra, B. Clifford. "Toward Loosely Coupled Programming on Petascale Systems", to appear, IEEE/ACM Supercomputing 2008.

**[Schmuck+2002]** F. Schmuck, R. Haskin, GPFS: A Shared-Disk File System for Large Computing Clusters, Proceedings of the USENIX FAST02 Conference on File and Storage Technologies, Monterey, California, 2002.

**[Thain+2005]** D. Thain, T. Tannenbaum, and M. Livny, "Distributed Computing in Practice: The Condor Experience" Concurrency and Computation: Practice and Experience, vol. 17, no. 2-4, pp. 323-356, Feb-Apr 2005.

**[Thain+2008]** D. Thain, C. Moretti, and J. Hemmes, Chirp: A Practical Global File system for Cluster and Grid Computing, Journal of Grid Computing, Springer, accepted for publication in 2008.

**[Thakur+1999]** R. Thakur, W. Gropp, E. Lusk. Data Sieving and Collective IO in ROMIO, 7th Symposium on the Frontiers of Massively Parallel Computation, 1999.

**[TUN]** Universal TUN/TAP Driver. http://vtun.sourceforge.net/tun

**[XAR]** XAR – eXtensible ARchiver Project home page, http://code.google.com/p/xar/

**[TOP500]** http://www.top500.org/system/9158

**[ZeptoOS]** The ZeptoOS Project. http://www.zeptoos.org/

**[Zhao+2007]** Y. Zhao, M. Hategan, B. Clifford, I. Foster, G. vonLaszewski, I. Raicu, T. Stef-Praun, M. Wilde, "Swift: Fast, Reliable, Loosely Coupled Parallel Computation" IEEE Workshop on Scientific Workflows 2007